\documentclass{article}
\usepackage{format}
\usepackage{latexsym}
\usepackage[english]{babel}
\usepackage{bm}
\usepackage[T1]{fontenc}
\usepackage[utf8]{inputenc}
\usepackage{ae}
\usepackage{graphicx}
\usepackage{subcaption}
\usepackage[graphicx]{realboxes}
\usepackage{epsfig}
\usepackage{color}
\usepackage{pgf}
\usepackage{pstricks,pst-grad}
\usepackage{float}
\usepackage{sidecap}
\sidecaptionvpos{figure}{c}
\usepackage[normalem]{ulem}
\usepackage{latexsym}
\usepackage{amsfonts}
\usepackage{amssymb}
\usepackage{amsthm}
\usepackage{amscd}
\usepackage{amsmath}
\usepackage{mathrsfs}
\usepackage{breqn}
\usepackage[bbgreekl]{mathbbol}
\usepackage{mathtools}
\usepackage{color}
\usepackage{colortbl}
\usepackage{framed}
\usepackage{tcolorbox}
\usepackage{verbatim}
\usepackage{pdfpages}
\usepackage{microtype}
\usepackage[inline]{enumitem}
\usepackage{verbatim}
\usepackage{placeins}
\usepackage[full]{textcomp}
\usepackage{wasysym}
\usepackage{appendix}
\usepackage[nolist,nohyperlinks]{acronym}
\usepackage{diagbox}
\usepackage{booktabs}
\usepackage{xr}
\usepackage{amsthm}
\usepackage{qcircuit}
\usepackage{physics}
\usepackage{comment}

\usepackage[backend=bibtex,sorting=none]{biblatex}
\usepackage{csquotes}
\usepackage{wrapfig}
\begin{document}

\title{Variational methods for Learning Multilevel Genetic Algorithms using the Kantorovich Monad}

\author{%
\textbf{Jonathan Warrell}$^{1,2,3*}$ \quad 
\textbf{Francesco Alesiani}$^{4*}$ \quad 
\textbf{Cameron Smith}$^{5,6*}$ \quad \\
\textbf{Anja M\"{o}sch}$^{4}$ \quad 
\textbf{Martin Renqiang Min}$^{1}$ \quad 
\\ \\
$^1$ NEC Laboratories America, Machine Learning Department,\\ Princeton, 08540, NJ, USA,\quad \\
$^2$ Department of Molecular Biophysics and Biochemistry, \\ Yale University, New Haven, 06520, CT, USA,\quad \\
$^3$ Program in Computational Biology and Bioinformatics, \\ Yale University, New Haven, 06520, CT, USA,\quad \\
$^4$ NEC Laboratories Europe, Kurfuerstenanlage
36, D-69115,\\  Heidelberg, Germany,\quad \\
$^5$ Molecular Pathology Unit and Center for Cancer Research, \\ Massachusetts General Hospital Research Institute, \quad \\
Department of Pathology, Harvard Medical School, Boston, MA, USA. \quad \\
$^6$ Broad Institute of MIT and Harvard, Cambridge, MA, USA, \quad \\
\texttt{*Equal Contribution. Corresponding author: jwarrell@nec-labs.org,}\\
\texttt{jonathan.warrell@yale.edu}\\
}

\maketitle

\begin{abstract}
Levels of selection and multilevel evolutionary processes are essential concepts in evolutionary theory, and yet there is a lack of common mathematical models for these core ideas.  Here, we propose a unified mathematical framework for formulating and optimizing multilevel evolutionary processes and genetic algorithms over arbitrarily many levels based on concepts from category theory and population genetics.  We formulate a multilevel version of the Wright-Fisher process using this approach, and we show that this model can be analyzed to clarify key features of multilevel selection.  Particularly, we derive an extended multilevel probabilistic version of Price's Equation via the Kantorovich Monad, and we use this to characterize regimes of parameter space within which selection acts antagonistically or cooperatively across levels.  Finally, we show how our framework can provide a unified setting for learning genetic algorithms (GAs), and we show how we can use a Variational Optimization and a multi-level analogue of coalescent analysis to fit multilevel GAs to simulated data.

\end{abstract}

\section{Introduction}

\begin{wrapfigure}[16]{R}{0.32\textwidth}
    \centering
    \includegraphics[width=0.25\textwidth]{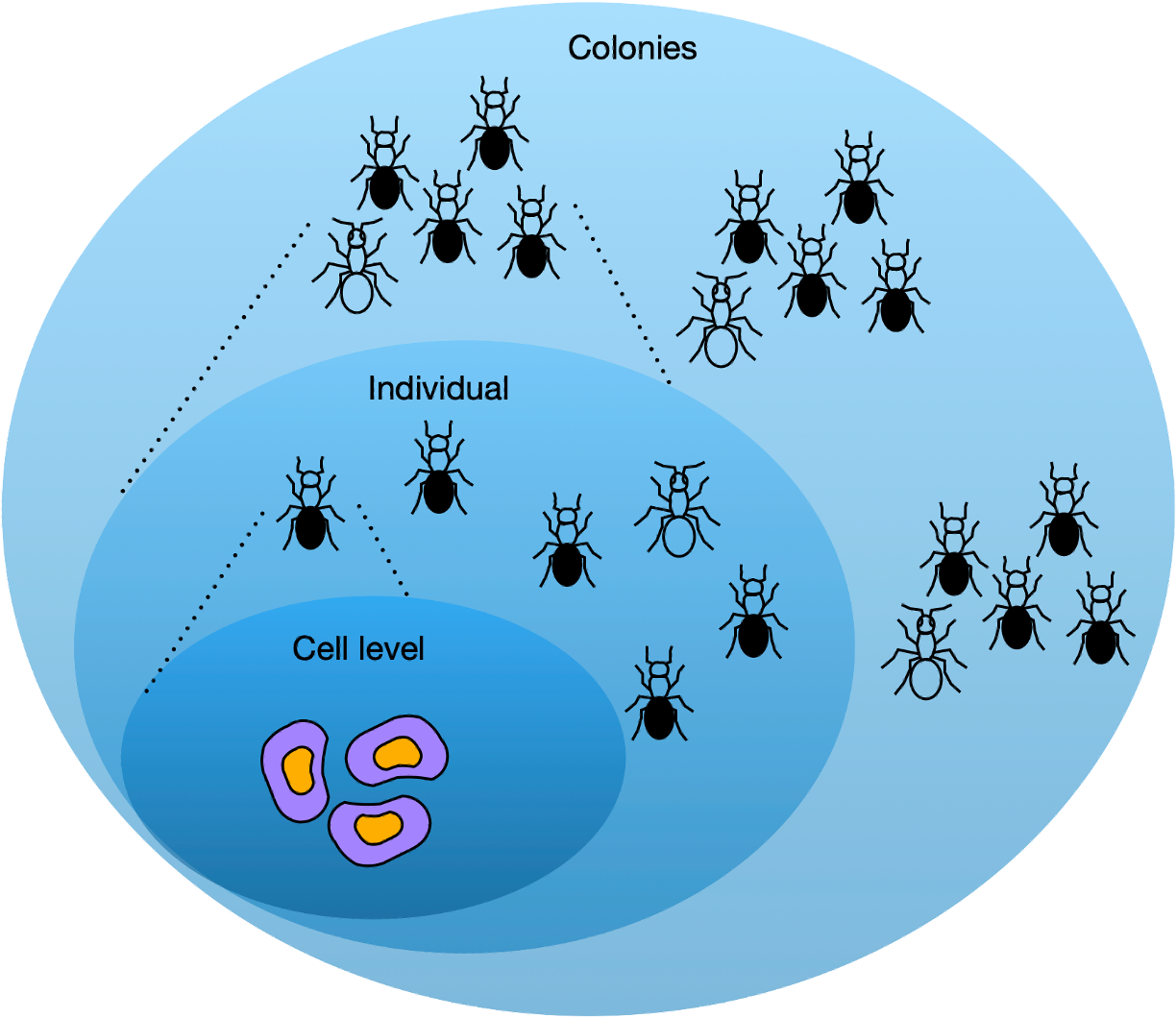}
    \caption{Multilevel evolutionary process for ant colonies.  Selection may act at the cellular, individual or colony level, and fitness any of these levels may conflict with fitness at any other.}
\end{wrapfigure}
An evolutionary process exhibits multilevel selection when it operates on units at multiple levels, each with its own definition of fitness.  For example, an evolutionary process involving ant colonies will exhibit units at least three levels: cells (basic units), ants (individuals), and colonies (groups of individuals). Such processes are important in diverse contexts.  Evolutionary theory hypothesizes that major transitions in evolution, such as the emergence of multicellularity, can be modeled in terms of multilevel selection \cite{calcott2011major,okasha2006evolution}.  At a different scale, cancer can also be viewed as a multilevel evolutionary process \cite{okasha2021cancer}.  Further, in the context of genetic algorithms, the use of hierarchical fitness functions has been shown to offer advantages when solving particular classes of problem \cite{watson2006compositional,watson1999hierarchically}. 
Despite the apparent ubiquity of multilevel evolutionary processes, few canonical models exist for studying this phenomenon in its essential aspects.  Prior models include evolutionary game theoretic models based on Moran dynamics \cite{traulsen2006evolution}, models based on hierarchical genetic algorithms \cite{watson2006compositional}, and group or kin selection models based on the recursive Price equation \cite{frank2020generalized,okasha2006evolution}. While the Moran dynamics approach of \cite{traulsen2006evolution} is informative as a minimal model (involving only cooperator and defector strategies), it is unclear how to extend this model to capture additional levels of important structure, such as large genotypes, complex maps between genotype and phenotype, and levels beyond the group level.  Further, while genetic algorithm based approaches such as \cite{watson2006compositional} allow a high level of generality, the connection to biological evolutionary models is indirect, limiting their use as a test-bed for studying such phenomena.  In addition, approaches based on the recursive Price equation \cite{frank2020generalized,okasha2006evolution} offer a general theoretical mechanism for probing phenomena related to multilevel selection; however, in general, they are not dynamically sufficient, and hence cannot be used as a complete model of an evolutionary process (for instance, for simulation).  Finally, while all of the above offer models of multilevel processes, they do not consider how the parameters of such models may be learned from data (real or synthetic) generated by such processes; indeed, this problem has received little attention, and it may be argued that developing methods of inference for multilevel processes is necessary for the objective identification and characterization of such processes in nature (for instance, determining the number of levels present, see \cite{warrell2020cyclic}).

In light of the above, we introduce here a minimal model of multilevel selection based on the Wright-Fisher model, which we refer to as a `multilevel Wright-Fisher process'.  Our model is minimal in the sense of including only those components necessary to embed a basic level of biological realism into the model; hence, we allow arbitrarily large genotypes and complex maps between genotype and phenotype to be embedded, while providing a consistent mechanism for extending the model to an arbitrary number of levels.  We are motivated here by a desire to both simulate and characterize such processes theoretically, as well as to use this model as a test-bed for developing algorithms to infer multilevel evolutionary processes in a sufficiently complex setting, which will have the required scalability to be applied to genomics scale data.  A particularly promising domain for such algorithms is in the area of cancer genomics, where the inference of predictive evolutionary models can help inform inference of tumor clonal structure and patient stratification for treatment selection, for example. 

Our approach is inspired by multiple recent advances in machine learning and computer science.  To allow maximal generality in our framework (in terms of number of levels and underlying spaces), we formulate our model in terms of stochastic processes over the Kantorovich Monad, which may be defined in the category of bounded metric spaces \cite{van2005metric}.  This allows the model to be extended over an indefinite number of levels by using the Wasserstein distance recursively to define similarity between pairs of members in a population, pairs of populations themselves, and pairs of meta-populations.  Conveniently, this approach also allows us to define recursive objective functions using the Wasserstein distance, which may be optimized efficiently.  Training objectives using the Wasserstein distance have been employed extensively in recent machine learning, as an alternative to likelihood-based objectives (see \cite{tolstikhin2017wasserstein}), but we are not aware of prior methods that have used the Kantorovich Monad as a means of defining recursive objectives for similarly hierarchically structured models.  

We provide a theoretical analysis of the properties of our framework, deriving a multilevel Price equation for the model, and using this to characterize regimes of parameter space within which selection acts antagonistically or cooperatively across levels.  In addition, we introduce a general optimization framework, combining Variational and Bayesian optimization methods with classical Coalescent analysis methods \cite{felsenstein2004inferring}.  There have been several recent attemps to use variational techniques to provide improved methods for phylogenetic inference \cite{moretti2021variational,zhang2018variational,zhang2020improved}.  Our methods offer distinct advantages over such approaches, by placing minimal constraints on the evolutionary processes to be learned, including being applicable to single and multilevel processes. Additionally, our approach is orthogonal to those above by applying a variational optimization (VO) as opposed to a variational inference (VI) approximation \cite{leordeanu2008smoothing}.

Sec. \ref{sec:model} outlines our multilevel Wright-Fisher model, and  Sec. \ref{sec:theory} then provides an analysis of theoretical properties of our framework. Sec. \ref{sec:training} introduces our optimization framework, including a multi-level analogue of coalescent analysis. Sec. \ref{sec:exp} illustrates our optimization approach on a small-scale example problem, involving learning a multilevel genetic algorithm over solutions to traveling salesman problems, and Sec. \ref{sec:disc} concludes with a discussion.

\section{Multilevel Evolutionary Processes on the Kantorovich Monad}\label{sec:model}

\begin{figure}[!t]
\centering
\includegraphics[width=0.70\linewidth]{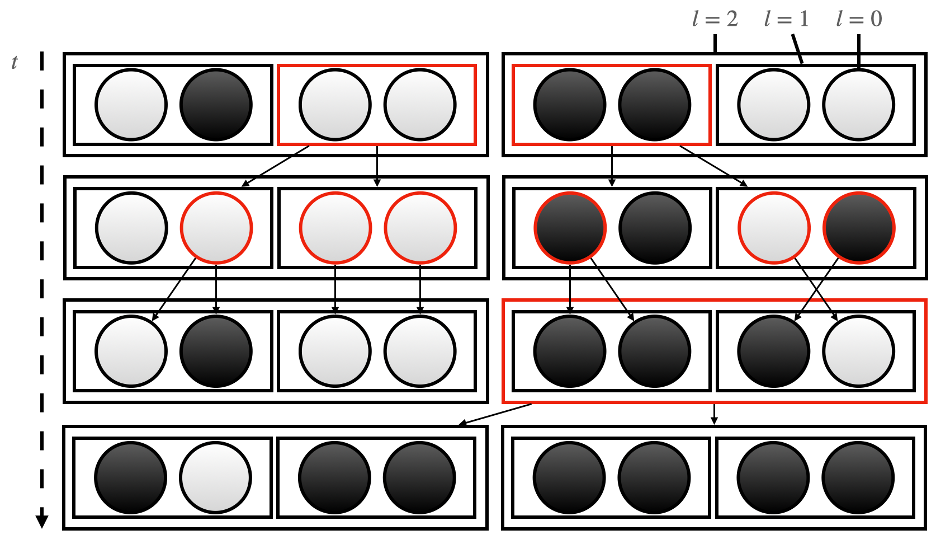}
\caption{Schematic of a Multi-level Evolutionary Process.  The population consists of $L=3$ levels, with two genotypes (open and closed circles) at level $l=0$ and 8 individuals, 4 groups of 2 individuals at level $l=1$, and 2 meta-groups of 2 groups (4 individuals) at level $l=2$.  Time $t$ flows down the page; at each time-step, a level is selected to reproduce, and the individuals (or groups) at that level which reproduce are highlighted in red.  Mutations may occur during reproduction; when a group at level $l$ reproduces, the Wasserstein distance at level $l$ determines the probability for the state of the offspring group.  The fitness of a group determines its probability of reproducing, and is determined by the fitness of the group members (here, the closed genotype has higher level 0 fitness than the open genotype), and the group's cohesion.}
\label{fig:fig1}
\end{figure}

Our framework is based on the Kantorivich Monad on the category of 1-bounded compact metric spaces and non-expansive maps ($\mathbb{KM}et_1$), as in \cite{van2005metric}.  Hence, we introduce the functor $\mathcal{B}:\mathbb{KM}et_1\rightarrow \mathbb{KM}et_1$, which takes a metric space $(X,d_1)$ to the metric space $(\mathcal{B}X,d_2)$, where $\mathcal{B}X$ is the set of Borel probability measures over $X$, and $d_2$ is the Wasserstein distance, defined as:

\begin{eqnarray}\label{eq_wasserstein}
d_2(\mu,\nu)=\min_{\tau\in T_{\mu,\nu}}\int_{X\times X} d_1(x_1,x_2) d\tau(x_1,x_2)
 \end{eqnarray} 

\noindent where is $T_{\mu,\nu}$ the set of Borel probability measures $\tau$ on $X \times X$, such that $\int_X \tau(x,.) d\mu(x) = \nu$ and $\int_X \tau(.,x) d\nu(x) = \mu$, and $\mu, \nu \in \mathcal{B}X$ are two given Borel probability measures.  $(\mathcal{B}f)(\mu)$ is defined as the push-forward measure.  Additionally, we require monad unit and multiplication functions.  The unit, $\eta_X:X\rightarrow \mathcal{B}X$ is defined as $\eta_X(x)=\delta(.;x)$, where $\delta(,;x)$ is the delta distribution at $x$, and the multiplication $\mu_X:\mathcal{BB}X \rightarrow \mathcal{B}X$ is defined as $(\mu_X \zeta)(B)=\int_X \nu(B) d\zeta(\nu)$, where $B\subset X$ is a Borel set.  With these definitions, $(\mathcal{B},\eta,\mu)$ can be shown to satisfy the monad properties \cite{van2005metric}.

We define a meta-population as a higher-order finite discrete distribution over a base space of genotypes (which themselves are not necessarily discrete).  Hence, we let $(X,d)$ denote our space of genotypes, where $d$ is an arbitrary distance function.  Then, a meta-population $x$ is parameterized by setting a level $l$, and a population size $n$.  A meta-population $x_{ln}$ is thus a member of $\mathcal{B}^L X$, having the form:

\begin{eqnarray}\label{eq_metapop}
x_{ln} = (1/n) \sum_{i = 1...n} \delta(.;x^{(i)}_{l-1,n})
 \end{eqnarray} 

\noindent where $x^{(i)}_{l-1,n}$, $i=1...n$, are $n$ meta-populations at level $l-1$, and we assume $l>0$.  We note that meta-populations at level $l=0$ are simply elements of $X$ (genotypes).

A multilevel Wright-Fisher process is defined by fixing a maximum level $L$ and a chosen population size $N$.  Additionally, we require a fitness function over the genotypes (at level-0), $f_0:X \rightarrow [0 \; 1]$, a cooperativity strength, $\kappa$, a mutation penalty $\lambda$, and potentially a base conditional distribution $\Delta_0(y|x)$ over the genotypes, representing the probability $x$ mutates to $y$.  The process then defines a Markov chain over meta-populations $x_{LN}$.  Hence, a sample from the process consists of a sequence $x^{(t=0)}_{LN},x^{(t=1)}_{LN},...,x^{(t=T)}_{LN}$.  To define the dynamics of the process, we first extend the fitness function across levels $1...L$.  For level $l$, we define the $f^l$, the fitness, as:

\begin{eqnarray}\label{eq_fitness}
f_l(x_{l}) = \kappa \cdot \psi(x_{l}) + \sum_{i = 1...N} f_{l-1}(x^{(i)}_{l-1})
 \end{eqnarray} 

\noindent where we drop the subscripts $n$ on the meta-populations for convenience (since these are fixed to $N$ at all levels, by definition), and the consistency function, $\psi$, is defined as:

\begin{eqnarray}\label{eq_cons}
\psi(x_{l}) = \frac{\sum_{i,j\leq N} (1-d_l(x^{(i)}_{l-1},x^{(j)}_{l-1}))}{N^2}.
 \end{eqnarray} 

The dynamics for the meta-population evolution then consists of, at each time step, sampling a level $l \in {0...L-1}$ to update (either uniformly, or according to a parameterized distribution $\{p_l|l=0...L-1\}$, and then updating the meta-population at level $l$ according to the following transition kernel:

\begin{eqnarray}\label{eq_trans}
P_l(x^{t}_{l}|x^{t-1}_{l}) \propto \sum_{\pi_l} \prod_{i=1...N^{L-l}} \tilde{f}(x^{t-1,\pi(i)}_{l}) 
\Delta_l(x^{t,i}|x^{t-1,\pi(i)}),
 \end{eqnarray} 

\noindent where:

\begin{eqnarray}\label{eq_fit2}
\tilde{f}(x^{t,i}_{l}) = \frac{f(x^{t,i}_{l})}{\sum_{j\in S_{i,l}}f(x^{t-1,j}_{l})} 
 \end{eqnarray} 

\noindent and $\pi_l$ denotes a mapping of the meta-populations at level $l$ and time $t$ to those at level $l$ and time $t-1$ (i.e., mapping a child meta-population or genotype to its parent at the previous time-step), which is restricted so that $\pi_l(i) \in S_i$, where $S_i$ is the set of meta-populations at level $l$ which are part of the same meta-population at level $l+1$ as $i$.  It remains to define $\Delta_l(.|.)$, the conditional mutation probability function.  We consider two possibilities here.  The first, which we use later in the synthetic experimental investigation, is to directly use the Wasserstein distance between meta-populations at level $l$ to define a mutation probability at level $l$; hence:

\begin{eqnarray}\label{eq_mut2}
\Delta_l(y_{l}|x_{l}) &\propto& \exp{-\lambda d_l(x_i,y_i)},
\end{eqnarray} 

\noindent While this is the simplest possibility, the constraint that $\Delta_l(.|.)$ be derived from the symmetric distance function $d_l$ at each level may not always be appropriate.  For this reason, we consider a second possibility, where a base distribution $\Delta_0(.|.)$ is provided independently, and for levels $l\geq 1$ we consider the number of `swaps' that have occurred between sub-populations at each level to generate a chosen meta-population from a given one.  For this case, a possible form for $\Delta_l(.|.)$ is:

\begin{eqnarray}\label{eq_mut2a}
\Delta_l(y_{l}|x_{l}) &\propto& \sum_{g_0,...,g_{l-1}} \prod_{l'=1...l-2} \exp{-\sum_{i \in 1...N^{l-l'+1}}\lambda
\left[ g_{l'}\left(\left\lfloor\frac{i}{N^{l-l'}} \right\rfloor\right)=\left\lfloor \frac{g_{l'-1}(i)}{N^{l-l'}} \right\rfloor\right]} \cdot \nonumber\\
&&\prod_{i=1...N^l} \Delta_0 (y_0^{(i)}|x_0^{(g_0(i))}),
\end{eqnarray} 

\noindent where $[.]$ is the Iverson bracket, and $g_{l'}$ is a permutation over the set $\{0...N^{l-l'}\}$.  

To complete the model specification, we assume an initial distribution over meta-populations at $t=0$, $P_0(x^{t=0}_{LN})$.  Finally, we note that when $L=1$, we recover the classical single-level Wright-Fisher process (with either a symmetric or asymmetric mutation matrix, corresponding to the two forms of $\Delta_l(.|.)$ discussed above).  Fig. \ref{fig:fig1} shows a schematic of a simple multi-level evolutionary process for illustration. 

\section{Theoretical Analysis: Conditions for conflict between levels of selection}\label{sec:theory}

For convenience, we analyse a simplified version of our model, in which there is no mutation, hence $p_m=0$, and the underlying genotype space ($\mathcal{X}$) is discrete.  First, we derive a multilevel version of the Price equation for our model.  This can be expressed as:

\noindent \textbf{Proposition 1 (Mixed Multilevel Price Equation):}  For a multilevel Wright-Fisher process with $p_m=0$, and a trait $\phi$ of individuals at level 0, we have that:
\begin{eqnarray}\label{eq_th1}
\mathbb{E}[\overline{\phi_{t+1}}-\overline{\phi_{t}}] = \sum_{l<L}p_l \cdot \text{Cov}(\phi_t,\hat{\Omega}^t_l),
 \end{eqnarray} 
\noindent where $\hat{\Omega}^t_{li} = N\tilde{f}(x^{t,\lfloor i/(N^{L-l}) \rfloor}_{ln})$, and $\tilde{f}$ is defined as in Eq. \ref{eq_fit2}.\\

\noindent \textit{Proof:} At time step $t$, a level $l<L$ is selected to reproduce, with probability $p_{l}$.  We wish to show that, conditional on level $l$ being selected, the expected change in $\bar{\phi}$ is $\text{Cov}(\phi_t,\hat{\Omega}^t_l)$, since, given this, the proposition follows directly.  For $l=0$, this follows immediately, since it is equivalent to the stochastic Price equation.  For $l>0$, consider individual $x^{ti}_{0,n}$ at level 0.  This belongs to meta-population $x^{t,\lfloor i/(N^{L-l}) \rfloor}_{ln}$ at level $l$.  When the latter reproduces, it copies each member of the meta-population exactly (since there is no mutation), and hence the expected number of copies of each member of the meta-population is identical to the expected number of copies of the group, $N\tilde{f}(x^{t,\lfloor i/(N^{L-l}) \rfloor}_{ln})$.  The expected (realized) fitness of $x^{ti}_{0,n}$ is thus $\hat{\Omega}^t_{li}$, and the expected change in $\phi$ conditioned on $l$ is $\text{Cov}(\phi_t,\hat{\Omega}^t_l)$ as required.
\qed \\

We note that Eq. \ref{eq_th1} is in fact a form of the Stochastic Price Equation, also termed Rice's Equation, as introduced in \cite{rice2004mathematical}.  Indeed, it can be derived as a special case of the recursive expansion of the Stochastic Price Equation (see \cite{simpson201110how}).  We term this special form a `Mixed Multilevel Price Equation' due its convenient expression as a probabilistic mixture over levels, resulting from the fact that the expectation is taken over the choice of which level reproduces at each time-step.  As we will see, this provides a minimal model for conflict between selection at multiple levels.  Further, we note that when $L=1$, and $\tilde{f}_{ti}=(w_{ti}/N^{L-l})$, where $w_{ti}$ is the realized fitness (i.e. the actual number of offspring of individual $i$ at time $t$), Eq. \ref{eq_th1} reduces to the deterministic Price equation with perfect transmission, $\mathbb{E}[\overline{\phi_{t+1}}-\overline{\phi_{t}}]=(1/\bar{w})\text{Cov}(\phi_t,w_t)$ (note that, since the population is a fixed size, $\bar{w}=1$).

We are particularly interested in analyzing the expected change in $f$ at different levels of the multilevel Wright-Fisher model, as the strength of the cooperativity potential ($\kappa$) is varied.  For a single-level model ($L=1$) with perfect transmission, the expected change in $f$ is always non-negative (a result analogous to Fisher's Fundamental Theorem, see \cite{okasha2006evolution}), since, setting $\phi=f$ in Eq. \ref{eq_th1}, we have:

\begin{eqnarray}\label{eq_th2}
\mathbb{E}[\overline{f_{t+1}}-\overline{f_{t}}] = \text{Cov}(f_t,\hat{\Omega}^t_l) = \frac{N}{\sum_i f_{ti}}\text{Var}(f_t) \geq 0,
 \end{eqnarray} 

\noindent In the multilevel Wright-Fisher model, an analogous result holds at all levels when $\kappa=0$; hence, the $f^l$ increases at all levels of the model when there is no cooperativity between the members of each meta-population:

\noindent \textbf{Proposition 2 (no conflict between levels when $\kappa=0$):}  For a multilevel Wright-Fisher process as above, when $\kappa=0$, we have that, for all levels $l=0...L-1$:

\begin{eqnarray}\label{eq_th3}
\mathbb{E}[\overline{f^l_{t+1}}-\overline{f^l_{t}}] \geq 0.
 \end{eqnarray} 
\\

\noindent \textit{Proof:}  First, we note that, since $\kappa=0$:

\begin{eqnarray}\label{eq_th3a}
f^l_{ti}=\sum_{j\in S^l_i} f^0_{tj},
\end{eqnarray} 

\noindent where $S_i^l=\{j|\lfloor j/(N^{L-l}) \rfloor=i\}$.  Hence, writing $\mathbb{E}[.|l']$ for the expectation under the assumption that level $l'$ reproduces, we have:

\begin{eqnarray}\label{eq_th3b}
\Delta^{l'}_t=\mathbb{E}[\overline{f^{l'}_{t+1}}-\overline{f^{l'}_{t}}|l'] = \sum_{j=1...N^{L-l'+1}}\frac{N}{\sum_{i\in S_j^{l'+1}} f^0_{ti}}\text{Var}(f^{l'}_{t,i\in S_j^1})\geq 0.
\end{eqnarray} 

\noindent Letting $F^l_t=\sum_i f^l_{ti}$, we have that $\overline{f^{l}_{t}}=(F^{l}_t/N^{L-l})$.  Further, we have that $F^l=F^{l'}$ for arbitrary $l$ and $l'$, since each $x^j_{0,n}$ appears in exactly one set $S^l_i$, and exactly one set $S^{l'}_{i'}$.  We thus have that:

\begin{eqnarray}\label{eq_th3c}
\overline{f^{l}_{t}}=(N^{L-l'}/N^{L-l})\overline{f^{l'}_{t}}=N^{l-l'}\overline{f^{l'}_{t}}.
\end{eqnarray} 

By summing across $l'$, and substituting Eq. \ref{eq_th3c} into Eq. \ref{eq_th3b}, we have:

\begin{eqnarray}\label{eq_th3d}
\mathbb{E}[\overline{f^{l}_{t+1}}-\overline{f^{l}_{t}}]=\sum_{l'} p_{l'} N^{l-l'} \Delta^{l'}_t \geq 0,
\end{eqnarray} 

\noindent and the proposition follows.
\qed \\

We now consider increasing $\kappa$ gradually.  As the following shows, for a value of $\kappa$ that can be specified in advance, there will necessarily be population states for each level $l<L-1$, where the expected change in $f^l$ is negative:

\noindent \textbf{Proposition 3 (conflict at all levels $l<L-1$ for $\kappa>\kappa^*$):}  We assume we have a multilevel Wright-Fisher process as above, where there exist at least two genotypes $x_1,x_2\in \mathcal{X}$ such that $f(x_1)\neq f(x_2)$, $p_L$ is uniform across levels, and (for convenience) $N=2h$ is even.  Then, there exists a value $\kappa^*$, such that when $\kappa>\kappa^*$, for each level $l<L-1$, there exists a meta-population state $x^{(l)}_{LN}$, such that:

\begin{eqnarray}\label{eq_th4}
\mathbb{E}[\overline{f^l(x^{(l)\dagger}_{LN})}-\overline{f^l(x^{(l)}_{LN})}] < 0.
 \end{eqnarray} 
\\

\noindent where $x^{(l)\dagger}_{LN}$ is the one time-step stochastic update of $x^{(l)}_{LN}$.

\noindent \textit{Proof:} We let $(x^*_1,x^*_2)\in \mathcal{X}^2$ be an arbitrary pair of genotypes, for which $f(x^*_1)>f(x^*_2)$.  To prove the proposition, we wish to find, for each level $l<L-1$, a meta-population state where the expected change in $\bar{f^l}$ is negative, for a sufficiently high setting of $\kappa$.  Hence, we fix a level $l$.  There are $N^{L-l}$ meta-populations at this level, $x_{ln}^{i=0...N^{L-l}-1}$.  Letting $S_i^l=\{j|\lfloor j/(N^{L-l}) \rfloor=i\}$, for level $l$ we consider the meta-population defined by fixing the genotypes at level 0 to:

\begin{eqnarray}\label{eq_th4a}
x_{0n}^i = \begin{cases}
  x^*_1  & \text{if} \;\; i\in \cup_{j=1...h}S_{N^{L-l}-j}^l \\
  x^*_2 & \text{otherwise},
  \end{cases}.
\end{eqnarray} 

Then, we consider the expected change in $\bar{f^l}$ when a given level $l'$ is selected for reproduction.  If $l'<l$, we have $\mathbb{E}[\bar{f}^l_{t+1}-\bar{f^l_{t}}|l'<l]=0$, since the members of each sub-population are all identical below level $l$, and hence have a Wasserstein distance of 0 and identical fitness, and thus the sub-population structure will be maintained exactly when this level reproduces.  

For $l'=l$ we consider reproduction within each of the $j=0...N^{L-l+1}-1$ groups at level $L+1$.  All of the meta-populations at level $l$ are uniform, and so their cooperativity term is 0.  Further, all of the reproducing groups are uniform except the last $h$.  Hence, the expected increase in $f^l$ from reproduction at $l$ is:

\begin{eqnarray}\label{eq_th4b}
\Delta^l_l=\mathbb{E}[\overline{f^{l}_{t+1}}-\overline{f^{l}_{t}}|l]=\frac{1}{N^{L-l+1}}\cdot\frac{N}{h(f^*_2+f^*_1)}\text{Var}([f^*_2, ... f^*_1, ...])=\frac{(f^*_2-\mu)^2+(f^*_1-\mu)^2}{N^{L-l}(f^*_2+f^*_1)}, \nonumber\\
\end{eqnarray} 

\noindent where $\mu=(f^*_2+f^*_1)/2$.  

For $l'>l$, we have that all meta-populations at level $l'$ are uniform, except for the last.  Letting $\mu_{l'}=((N^{l'}-h)f^*_2+hf^*_1)/N^{l'}$, we thus have:

\begin{eqnarray}\label{eq_th4c}
f^{l',i}_{l'n} = \begin{cases}
  f^{l',*}_{1} = N^{l'}\mu_{l'}+\kappa(1-\frac{(N-1)^2+1}{N^2})(1-d_l(x^{l'-1,*}_1,x^{l'-1,*}_{N^{L-l'-1}-1}))  & \text{if} \;\; i = N^{L-l'}-1\\
  f^{l',*}_{2} = N^{l'}f^*_2 & \text{otherwise},
  \end{cases}. 
\end{eqnarray}

The expected change in fitness at level $l$ given reproduction occurs at level $l'>l$ is therefore:

\begin{eqnarray}\label{eq_th4d}
\Delta^{l'}_l=\mathbb{E}[\overline{f^{l}_{t+1}}-\overline{f^{l}_{t}}|l']=\frac{1}{N^{L-l'+1}}\cdot\left( \frac{N((N-1)f^{l',*}_2 f^*_2 + f^{l',*}_1 f^*_1)}{(N-1)f^{l',*}_2+f^{l',*}_1} - \mu \right).
\end{eqnarray} 

We note that, while $f^*_1>f^*_2$, we can make $f^{l',*}_2-f^{l',*}_1$ arbitrarily large by increasing $\kappa$ in Eq. \ref{eq_th4c}.  Note, that the maximum increase in the number of $x^*_1$'s when level $l$ reproduces is $N^l/2$, while the maximum decrease in the number of $x^*_1$'s when level $l'$ reproduces is also $N^l/2$.  Hence, the maximum value of $\Delta^l_l$ is $((N^l/2)\mu)/(N^l)=\mu/2$, and similarly, the maximum value of $\Delta^{l'}_l$ is $-\mu/2$.  However, since $f^*_1,f^*_2<\infty$, $\Delta^l_l<\mu/2$, meaning that we can find a value $\kappa_{l,l'}$ to ensure that $|\Delta^{l'}_l|>\Delta^{l}_l$.  We thus set:

\begin{eqnarray}\label{eq_th4e}
\kappa^*=\max_{(l,l')\in\{1...L\}^2,l'>l} \kappa_{l,l'}
\end{eqnarray} 

\noindent and the proposition follows.
\qed \\

Prop. 3 shows that, for a sufficiently high $\kappa$, $f$ may decrease at all levels of a multilevel Wright-Fisher process, except for the highest level $L-1$ (note that reproduction does not occur at level $L$ itself, since there is a single instance at this level, representing the global meta-population).  We may wonder, however, whether for certain values of $\kappa$, we are guaranteed to find population states with negative expected change in $f^{L-1}$.  The following example answers this in the negative:

\noindent \textbf{Example 1 (model with non-decreasing $f^{L-1}$ for all states):} Consider a multilevel Wright-Fisher process defined over a genotype space of two genotypes, $\mathcal{X}=\{x^*_1,x^*_2\}$, with two levels, $L=2$, and $N=2$.  Let the fitness function for level $0$ have values $f^{*}_1$ and $f^{*}_2$ for the genotypes respectively, where $f^{0,*}_1>f^{0,*}_2$, and let their distance at level 0 be $2d^*$.  The meta-population states may thus be represented $[0,0|0,0]$, $[0,0|0,1]$ ... $[1,1|1,1]$, where $1$ indicates an individual with genotype $x^*_1$.  There are thus 16 possible meta-population states. 

We consider the expected change in $f^1$ given level $0$ is chosen to reproduce (by definition, if level 1 reproduces, $f^1$ is expected to be be non-negative).  For states $[0,0|0,0]$ and $[1,1|1,1]$, $[0,0|1,1]$ and $[1,1|0,0]$, the expected change in $f^1$ when level 0 reproduces is 0.  For all other states, there exists at least one $[0,1]$ or $[1,0]$ sub-population.  For each such sub-population, when level 0 reproduces, the probabilities of the subpopulations $[0,0],[0,1] \text{or} [1,0],[1,1]$ are proportional to $(f^{*}_2)^2$, $2f^{*}_1f^{*}_2$ and $(f^{*}_1)^2$ respectively.  Hence, to ensure an expected increase in $f^1$, we require:

\begin{eqnarray}\label{eq_ex1}
(\tilde{f}^*_2)^2(2f^*_2+\kappa)+(\tilde{f}^*_1)^2(2f^*_1+\kappa)+2\tilde{f}^*_1\tilde{f}^*_2(f^*_1+f^*_2+\kappa(1-d^*)) &>& (f^*_1+f^*_2+\kappa(1-d^*)) \nonumber \\
(\tilde{f}^*_2)^2(2f^*_2+\kappa)+(\tilde{f}^*_1)^2(2f^*_1+\kappa) &>& (1-2\tilde{f}^*_1\tilde{f}^*_2)(f^*_1+f^*_2+ \nonumber \\
&& \kappa(1-d^*)) 
\end{eqnarray} 

\noindent Hence, rearranging, we have:

\begin{eqnarray}\label{eq_ex1a}
d^* &>& \frac{(1-2\tilde{f}^*_1\tilde{f}^*_2)(f^*_1+f^*_2+\kappa)-(\tilde{f}^*_2)^2(2f^*_2+\kappa)-(\tilde{f}^*_1)^2(2f^*_1+\kappa)}{(1-2\tilde{f}^*_1\tilde{f}^*_2)\kappa} \nonumber \\
&=& \frac{(1-2(\tilde{f}^*_1)^2-2\tilde{f}^*_1\tilde{f}^*_2)f^*_1 + (1-2(\tilde{f}^*_2)^2-2\tilde{f}^*_1\tilde{f}^*_2)f^*_2 
+ (1-(\tilde{f}^*_2)^2-(\tilde{f}^*_1)^2-2\tilde{f}^*_1\tilde{f}^*_2)\kappa}{(1-2\tilde{f}^*_1\tilde{f}^*_2)\kappa}  \nonumber \\
&=& \frac{(1-2\tilde{f}^*_1)f^*_1 + (1-2\tilde{f}^*_2)f^*_2}{(1-2\tilde{f}^*_1\tilde{f}^*_2)\kappa}
= \frac{(\tilde{f}^*_2-\tilde{f}^*_1)f^*_1 + (\tilde{f}^*_1-\tilde{f}^*_2)f^*_2}{(1-2\tilde{f}^*_1\tilde{f}^*_2)\kappa} 
= \frac{(\tilde{f}^*_1-\tilde{f}^*_2)(f^*_2-f^*_1)}{(\tilde{f}^*_1-\tilde{f}^*_2(\tilde{f}^*_1-\tilde{f}^*_2))\kappa}.
\end{eqnarray} 

\noindent Since $f^*_2-f^*_1<0$, the RHS of Eq. \ref{eq_ex1a} is negative, meaning that the change in $f^1$ is non-decreasing for any choice of $d^*$ and $\kappa$ for all states.
\qed \\

The following example. however, shows that it is possible to construct models in which states exist with an expected decrease in $f^{L-1}$ for certain states.

\noindent \textbf{Example 2 (model with decreasing $f^{L-1}$ for some states):}  We consider a model as in example 1, with $N>2$, and focus on the population state $[0,0,...,1|0,0,...,1|...]$.  When level 1 reproduces, since all sub-populations are identical, the expected change in $f^1$ is 0.  When level 0 reproduces, the expected change in $f^1$ is as follows:

\begin{eqnarray}\label{eq_ex2}
\mathbb{E}[\bar{f}^1_{t+1}-\bar{f}^1_{t}|l=0] &=& \left(\sum_m p_m F_m\right) - F_1 \nonumber \\
&=& \mathbb{E}[\bar{f}^0_{t+1}-\bar{f}^0_{t}|l=0] + \kappa\left(\left(\sum_m p_m D_m\right) - D_1\right)
\end{eqnarray} 

\noindent where $F_m$ is the fitness at level 1 of a sub-population with $m$ 1's, $p_m$ is the probability of sampling such a sub-population at time $t+1$ from the sub-population $[0,0,...,0,1]$ at time $t$, and $D_m$ is the average distance between a pair of elements sampled with replacement from such a sub-population.  Hence, we have:

\begin{eqnarray}\label{eq_ex2a}
p_m &=& \binom{N}{m}(\tilde{f}^*_1)^m(\tilde{f}^*_2)^{N-m} \nonumber \\
D_m &=& 1 - \frac{m(N-m)d^*}{N^2} \nonumber \\
F_m &=& m\tilde{f}^*_1 + (N-m)\tilde{f}^*_2 + \kappa D_m
\end{eqnarray} 

Since the term $\mathbb{E}[\bar{f}^0_{t+1}-\bar{f}^0_{t}|l=0]$ on the RHS of Eq. \ref{eq_ex2} is necessarily positive, the LHS will be negative only if $(\sum_m p_m D_m) - D_1$ is negative, and $\kappa$ is sufficiently large.  We perform a computational search over models parameterized as above, and find that, while this condition is not satisfied for models with $N<6$, for $N=5$ the condition is satisfied for all $d^*\in\{0.01,0.02,...,0.99\}$ and $\tilde{f}^*_1 \in \{0.51,0.52,...,0.72\}$ (note that $\tilde{f}^*_2=1-\tilde{f}^*_1$).  We similarly found solutions for all values $N>5$ tested.
\qed \\

We conclude that, while conflict between levels of selection necessarily exists for sufficiently high $\kappa$ for all levels $l<L-1$, for level $L-1$ the existence of such conflict depends on the specific parameters of the process.

\section{Variational Methods for Learning Multilevel Processes}\label{sec:training}

We outline below three methods for fitting the parameters of a multilevel evolutionary process, based on Variational Optimization (VO) \cite{leordeanu2008smoothing}, Simultaneous Perturbation Stochastic Perturbation (SPSA) \cite{spall2001stochastic}, and Monte-Carlo Expectation Maximization (MC-EM) \cite{ruth2024review}.  Each of these methods may be used by combining both forward simulations of our model and coalescent (backwards) simulations, which we introduce below.  In addition, VO and MC-EM are both variational methods, while SPSA is a stochastic gradient descent approach; further, while MC-EM operates on a traditional Evidence Lower-Bound (ELBO) objective, VO and SPSA directly optimize a multi-level Wasserstein distance objective, which conveniently represents the multi-level structure of our problem.  Comparing the performance of these approaches in Sec. \ref{sec:exp} thus allows us to compare the relative advantages of using variational/gradient-based methods, ELBO/Wasserstein objectives, and forward/coalescent simulations.

Throughout this section, we denote the parameters collectively as $\theta$ (with dimension $D_\theta$, which may include the parameters of $f$ and $\Delta$, along with $\kappa$ and $\lambda$).  In each case, for simplicity we assume we have access to one meta-population generated by the ground-truth process at a chosen time-point, $T$, which is denoted $\mathbf{x}^{T\dagger}_L$.

\subsection{Variational Optimization (VO) using forward simulations}\label{sec:vo1}

Following \cite{leordeanu2008smoothing}, we introduce a variational distribution over the parameters $\theta$, which we take to be a Gaussian with a symmetric covariance matrix.  At a given meta-epoch $\tau$, this variational distribution has the form:

\begin{eqnarray}\label{eq_var}
\theta_\tau \sim \mathcal{N}(.|\mu_\tau,\sigma_\tau I),
\end{eqnarray} 

\noindent where $\mu_\tau$ is a vector of mean values, $\sigma_\tau$ is a scalar, and $I$ is the identity matrix.  At meta-epoch $\tau$, we draw $S$ samples from Eq. \ref{eq_var}, $\theta^1_\tau...\theta^S_\tau$, and for each, we run $R$ forward multilevel Wright-Fisher processes to generate meta-population samples $\mathbf{x}^{T,s,r}_L$, for $r=1...R$.  We then use these to calculate the following score:

%% could we use exp - d ?
\begin{eqnarray}\label{eq_var1}
F_s^\tau = 1-\mathbb{E}_{\mathbf{x}^{T}_L\sim P^{L,T}_{\theta_\tau^s}}[d_L(\mathbf{x}^{T}_L,\mathbf{x}^{T\dagger}_L)] \approx 1 - \frac{1}{R} \sum_r d_L(\mathbf{x}^{T,s,r}_L,\mathbf{x}^{T\dagger}_L),
\end{eqnarray} 

\noindent where $d_L(.,.)$ is the Wasserstein distance at level $L$, and $P^{L,T}_\theta$ is the distribution over $\mathcal{B}^L X$ at time-step $T$ induced by the multilevel Wright-Fisher process with parameters $\theta$.

We then apply the smoothing-based optimization (SMO) updates from \cite{leordeanu2008smoothing}, to update $\mu$ and $\sigma$:

\begin{eqnarray}\label{eq_var2}
\mu_{\tau+1} &=& \frac{\sum_s F_s \theta^s_\tau}{\sum_s F_s} 
 \end{eqnarray} 
\begin{eqnarray}\label{eq_var2b}
\sigma_{\tau+1} &=& \sqrt{\frac{\sum_s F_s |\theta^s_\tau-\mu_{\tau}|^2_2}{D_\theta \sum_s F_s}}.
 \end{eqnarray} 

\noindent The updates in Eq. \ref{eq_var2} and \ref{eq_var2b} can be shown to improve the value of $F$ in expectation as $S\rightarrow \infty$ and $R\rightarrow \infty$ (see \cite{leordeanu2008smoothing}), and hence reduce the expected Wasserstein distance between the simulated and ground truth processes:

\begin{eqnarray}\label{eq_var3}
\mathbb{E}_{\theta\sim Q_{\tau+1}}\mathbb{E}_{\mathbf{x}^{T}_L\sim P^{L,T}_\theta}[d_L(\mathbf{x}^{T}_L,\mathbf{x}^{T\dagger}_L)]\leq \mathbb{E}_{\theta\sim Q_\tau}\mathbb{E}_{\mathbf{x}^{T}_L\sim P^{L,T}_\theta}[d_L(\mathbf{x}^{T}_L,\mathbf{x}^{T\dagger}_L)],
 \end{eqnarray} 

\noindent where $Q_\tau = \mathcal{N}(.|\mu_\tau,\sigma_\tau I)$.

\subsection{Variational Optimization (VO) using forward and coalescent simulations}\label{sec:vo2}

Exclusively using forward simulations, as in Sec. \ref{sec:vo1}, may lead to noisy estimates of the expected Wasserstein distance for a given sample $\theta_s$, due to the small likelihood of generating samples close to the ground-truth data.  To stabilize the expected Wasserstein distance estimates, we assume we have access to a coalescent sampler $\mathcal{C}(.|\mathbf{x}^{T\dagger}_L,\theta)$, which generates possible meta-population trajectories given a final ground-truth state.  We then consider the following expression for $F^\tau_s$ in Eq. \ref{eq_var1}:

\begin{eqnarray}\label{eq_coalesc}
F^\tau_s = 1 - (1-P^{L,T}_{\theta_\tau^s}(d_L(\mathbf{x}^{T}_L,\mathbf{x}^{T\dagger}_L)=0))\mathbb{E}_{\mathbf{x}^{T}_L\sim P^{L,T}_{\theta_\tau^s}}[d_L(\mathbf{x}^{T}_L,\mathbf{x}^{T\dagger}_L)|d_L(\mathbf{x}^{T}_L,\mathbf{x}^{T\dagger}_L)>0].
\end{eqnarray} 

\noindent Since simulations generated by the forward sampler are highly likely to have non-zero Wasserstein distance to the ground-truth, we may use rejection sampling to estimate the conditional expectation in the last RHS term of Eq. \ref{eq_coalesc}, and we assume that this can be done through generating $R$ samples $\mathbf{x}^{T,s,r}_L$ as in Eq. \ref{eq_var1}, with the added non-zero Wasserstein distance constraint.  To estimate the scaling factor, $P^{L,T}_{\theta_\tau^s}(d_L(\mathbf{x}^{T}_L,\mathbf{x}^{T\dagger}_L)=0)$, we may use a coalescent sampler, since by definition, all samples from $\mathcal{C}(.|\mathbf{x}^{T\dagger}_L,\theta)$ have zero Wasserstein distance to the ground-truth.  We thus generate $R$ further coalescent samples, $\tilde{\mathbf{x}}^{s,r}_L$.  Assuming we can straightforwardly calculate the probability of each sample under the forward $\mathcal{P}$ and coalescent $\mathcal{C}$ distributions, we may thus estimate the scaling factor using importance sampling:

\begin{eqnarray}\label{eq_coalesc2}
P^{L,T}_{\theta_\tau^s}(d_L(\mathbf{x}^{T}_L,\mathbf{x}^{T\dagger}_L)=0) &=& \frac{Z^0_{\mathcal{P}_{\theta_\tau^s}}}{Z_{\mathcal{C}_{\theta_\tau^s}}} \nonumber\\
&=& \mathbb{E}_{\mathbf{x}_L\sim\mathcal{C}_{\theta_\tau^s}}\left[\frac{\mathcal{P}_{\theta_\tau^s}(\mathbf{x}_L)}{\mathcal{C}_{\theta_\tau^s}(\mathbf{x}_L))}\right] \nonumber \\
&\approx& \frac{1}{R}\sum_r \exp \left(\log \mathcal{P}_{\theta_\tau^s}(\tilde{\mathbf{x}}^{s,r}_L) - \log \mathcal{C}_{\theta_\tau^s}(\tilde{\mathbf{x}}^{s,r}_L)\right),
\end{eqnarray} 

\noindent where $Z^0_{\mathcal{P}_{\theta_\tau^s}}$ is the normalization factor for the forward distribution conditioned on observing zero Wasserstein distance with the ground-truth, and $Z_{\mathcal{C}_{\theta_\tau^s}}$ is the normalization factor for the coalescent distribution (which we assume to be 1, since we can straightforwardly calculate the probability of each sample under $\mathcal{C}$).

\noindent \textbf{Multilevel Coalescent Process:} We define a multilevel coalescent process $\mathcal{C}(.|\mathbf{x}^{T\dagger}_L,\theta)$ as follows. For each time-step in descending order, $t=(T-1)...0$, we first choose a focal level $l\in\{0...L-1\}$.  We then cycle through the $N^{L-l}$ meta-populations at level $l$, at time $t+1$.  For each, we assign a parent meta-population at time $t$, independently.  We then attempt to assign a state to each of the parent populations, $x_l^{t,j}$.  If $x_l^{t,j}$ has one or more children, one of the children $x_l^{t+1,k}$ is chosen at random.  Then, the state of the parent is chosen by sampling $x_l^{t,j}\sim \Delta_l(.|x_l^{t+1,k})$. However, if $x_l^{t,j}$ has no children, its state is left undefined.  Further, we note that not all of the children may have a defined state; if only some children have defined states, $x_l^{t+1,k}$ is restricted to these cases, and in cases where the child's state is itself not fully defined (for instance, if only some of the members of a meta-population have been defined), the operator $\Delta_l(.|x_l^{t+1,k})$ must be extended to condition on partial states (in the case of Eq. \ref{eq_mut2}, which we use in the experimentation, this is straightforward, since we may simply use as many individuals as are defined at each level to calculate the multi-level Wasserstein distance, while preserving the number of defined individuals at each level between  $x_l^{t+1,k}$ and $x_l^{t,j}$).  Finally, when the coalescent sampler has reached time $t=0$, any undefined $l=0$ individuals at $t=0$ have their states sampled from the initial distribution, and remaining undefined states at $t>0$ are sampled by forward sampling the multilevel Wright-Fisher process with parameters $\theta$, while maintaining the level assignments at each time-step used in the backward sampling phase, and conditioning on all those individuals whose states have already been assigned in the backwards pass.

\subsection{Simultaneous Perturbation Stochastic Perturbation (SPSA)}\label{sec:spsa}

Simultaneous Perturbation Stochastic Perturbation (SPSA) \cite{spall2001stochastic} provides a method for performing stochastic gradient descent on a function which is easy to evaluate, but for which an analytic gradient is not available. The method requires hyperparameters $a,A,\alpha,c,\gamma$, and for epoch $\tau$ we define $a_\tau = (a/(\tau+A))^\alpha$ and $c_\tau=(c/\tau)^\gamma$. Then, $\theta$ is initialized randomly at epoch 0, and for each epoch $\tau$ we sample a vector $\delta \in \{-1,1\}^{D_\theta}$, representing randomized perturbations to the current value of $\theta$ (where $\delta(i)$ is sampled independently and uniformly from $\{-1,1\}$).  Assuming we wish to minimize the function $F(\theta)$, we then use the following gradient estimator and update at each epoch:

\begin{eqnarray}\label{eq_spsa}
\hat{g}_\tau &=& \frac{F(\theta_\tau+c_\tau \delta)-F(\theta_\tau-c_\tau \delta)}{2c_\tau} \nonumber \\
\theta_{\tau+1} &=& \theta_\tau - a_\tau \hat{g}_\tau.
\end{eqnarray} 

To train our model, we use the expected Wasserstein distance as our function of interest, which we approximate as in Eq. \ref{eq_var1} using $R$ Monte-Carlo samples:

\begin{eqnarray}\label{eq_spsa1}
F(\theta) = \mathbb{E}_{\mathbf{x}^{T}_L\sim P^{L,T}_{\theta}}[d_L(\mathbf{x}^{T}_L,\mathbf{x}^{T\dagger}_L)] \approx \frac{1}{R} \sum_r d_L(\mathbf{x}^{T,r}_L,\mathbf{x}^{T\dagger}_L).
\end{eqnarray} 

Further, as in Eq. \ref{eq_coalesc}, we may incorporate coalescent simulations to provide an alternative estimator of $F(\theta)$ based on the following equivalent expression for $F(\theta)$:

\begin{eqnarray}\label{eq_spsa2}
F(\theta) = (1-P^{L,T}_{\theta}(d_L(\mathbf{x}^{T}_L,\mathbf{x}^{T\dagger}_L)=0))\mathbb{E}_{\mathbf{x}^{T}_L\sim P^{L,T}_{\theta}}[d_L(\mathbf{x}^{T}_L,\mathbf{x}^{T\dagger}_L)|d_L(\mathbf{x}^{T}_L,\mathbf{x}^{T\dagger}_L)>0].
\end{eqnarray} 

\subsection{Monte-Carlo Expectation Maximization (MC-EM)}\label{sec:mcem}

We desire to estimate $\theta$, given a ground-truth sample $\mathbf{x}^{T\dagger}_L$ representing a final meta-population at time-step $T$.  Hence, we may treat the meta-population states at time-steps $t<T$ as latent variables, and use an EM-style approach to train $\theta$ by alternately estimating the posterior over $\mathbf{x}^{0...T-1}_L$ and updating our estimate of $\theta$ conditional on this posterior.  In a Monte-Carlo Expectation Maximization approach (MC-EM, see \cite{ruth2024review}), the posterior over $\mathbf{x}^{0...T-1}_L$ is represented by sampling.  In our case, since we cannot directly sample from the required posterior, we use importance sampling to represent it, by performing multiple forward and/or coalescent simulations using the current estimate for $\theta$.

Hence, at epoch $\tau$, for the forward sampling case, we run $S$ forward simulations by sampling from $P^{L,0...T}_{\theta_\tau}$, where $\theta_\tau$ is the estimate for $\theta$ at epoch $\tau$.  We denote by $\mathbf{x}^{s}_L$ the $s$'th simulation, where the final time-step of each simulation has been replaced by the ground-truth state $\mathbf{x}^{T\dagger}_L$ (while maintaining all ancestral relationships in the generated sample).  During the M-step, $\theta$ is updated by optimizing the following importance sampling estimate for the ELBO objective:

\begin{eqnarray}\label{eq_mcem}
L_1(\theta_{\tau+1}) &=& \sum_s \frac{P^{L}_{\theta_\tau}(\mathbf{x}^{s}_L)}{P^{L,0...T-1}_{\theta_\tau}(\mathbf{x}^{s}_L)\cdot K_1} \log P^{L}_{\tau+1}(\mathbf{x}^{s}_L) \nonumber \\
K_1 &=& \sum_s \frac{P^{L}_{\theta_\tau}(\mathbf{x}^{s}_L)}{P^{L,0...T-1}_{\theta_\tau}(\mathbf{x}^{s}_L)}.
\end{eqnarray} 

To incorporate coalescent simulations into the MC-EM optimizer, we draw $s=1...S$ forward and $r=1...R$ coalescent simulations ($\tilde{\mathbf{x}}^{r}_L$ )at each epoch, and use the following Monte-Carlo estimator for the ELBO bound:

\begin{eqnarray}\label{eq_mcem1}
L_2(\theta_{\tau+1}) &=& \frac{L_1}{2} + \sum_r \frac{P^{L}_{\theta_\tau}(\tilde{\mathbf{x}}^{r}_L)}{2\cdot\mathcal{C}_{\theta_\tau}(\tilde{\mathbf{x}}^{r}_L)\cdot K_2} \log P^{L}_{\tau+1}(\tilde{\mathbf{x}}^{r}_L) \nonumber \\
K_2 &=& \sum_r \frac{P^{L}_{\theta_\tau}(\tilde{\mathbf{x}}^{r}_L)}{\mathcal{C}_{\theta_\tau}(\tilde{\mathbf{x}}^{r}_L)}.
\end{eqnarray} 

\section{Results on a synthetic dataset}\label{sec:exp}

We design a synthetic problem to test our optimization approach, based on learning the parameters of a fitness function for a multi-level evolutionary process, whose genotypes represent solutions to a Traveling Salesman Problem (TSP).

The underlying TSP consists of finding the shortest path connecting $M$ cities, whose coordinates sampled from a standard 2d Gaussian distribution.  Since the coordinates of the cities implicitly define the fitness of a chosen path, we denote the $m$'th city's coordinates as $\theta_{m}=(\theta_{m,1},\theta_{m,2})$, and our goal in training the GA is to learn $\theta$, hence the coordinates of the cities.  Further, we denote the distance between cities $m_1$ and $m_2$ as $d_{\text{TSP}}(m_1,m_2)$.  The underlying space of genotypes $X$ is thus the set of permutations on $M$ elements, and the (level 0) fitness function of genotype $x\in\mathcal{X}$ is defined as:
\begin{eqnarray}\label{eq:fit}
f_0(x)=\exp(-\beta \cdot \sum_{i=1...M-1}d_{\text{TSP}}(x_i,x_{i+1}))
\end{eqnarray} 
\noindent where $\beta$ is an inverse temperature parameter.  A simulation of the GA consists of sampling a meta-population of genotypes of size $N^L$ uniformly from $X$, and updating these for $T$ time-steps using the multilevel update rule (Eq. \ref{eq_trans}). We use a mutation function $\Delta'_l(.|.)$ which is a variation of Eq. \ref{eq_mut2}, defined as:
\begin{eqnarray}\label{eq:delt}
\Delta'_l(x|y)=(1-p_m)[x=y]+p_m[x\neq y]\frac{\Delta_l(x|y)}{1-\Delta_l(y|y)}
\end{eqnarray} 
\noindent where $p_m$ denotes the probability of a mutation occurring, and $[.]$ is the Iverson bracket.

We run $N_{\text{data}}=5$ simulations as above up to time-step $T$, and a dataset is generated by including only the final population state from each simulation.  Our ground-truth training set may thus be represented as $X\in \{1...M\}^{LNM}$, where $X_{lnm}$ represents the city ($1...M$) visited by the $n$'th population member in simulation $l$ at the $m$'th location on the route.  For each simulation, we sample a new ground-truth $\theta$ vector, and for all simulations we fix the values of $L=3, M=5, N=4, T=5, \beta=2, \kappa=1, \lambda=5, p_m=0.25$, which we treat as known.  

We compare the performance of each of the optimization algorithms discussed in Sec. \ref{sec:training} (VO, SPSA and MC-EM), comparing in each case the performance when the algorithm is restricted to use only forward simulations against the performance when both forward and coalescent simulations are used.  Since both VO and SPSA approaches minimize the expected multi-level Wasserstein distance to the observed data, we compare the best Wasserstein distances achieved by these methods as a performance metric, while for the MC-EM approaches, we use the estimated ELBO bound as a metric.  In addition, for all methods we consider the accuracy with which the method infers the positioning of the cities in the TSP problem (which implicitly define the fitness function) for all methods; hence, for the best performing model, we consider the minimum Euclidean Distance achieved according to the best alignment of the city positions predicted by the method and the ground-truth positions, subject to an arbitrary rotation and displacement.  For the VO forward approach, we use $S=5$ samples for $\theta$ per epoch and $R=6$ forward simulations per sample, and for the VO coalescent approach we set $S=5$ and $R=3$ (the latter generating 6 forward and coalescent simulations combined).  For SPSA, we use $R=6$ samples for the forward simulation model, and $R=3$ forward and coalescent simulations each for the mixed model.  Similarly, for MC-EM, we use $S=6$ forward samples for the forward model, and $S=R=3$ forward and coalescent simulations respectively for the mixed model.  We train each algorithm for 10 epochs.

\begin{figure}[!t]
\centering
\includegraphics[width=0.8\linewidth]{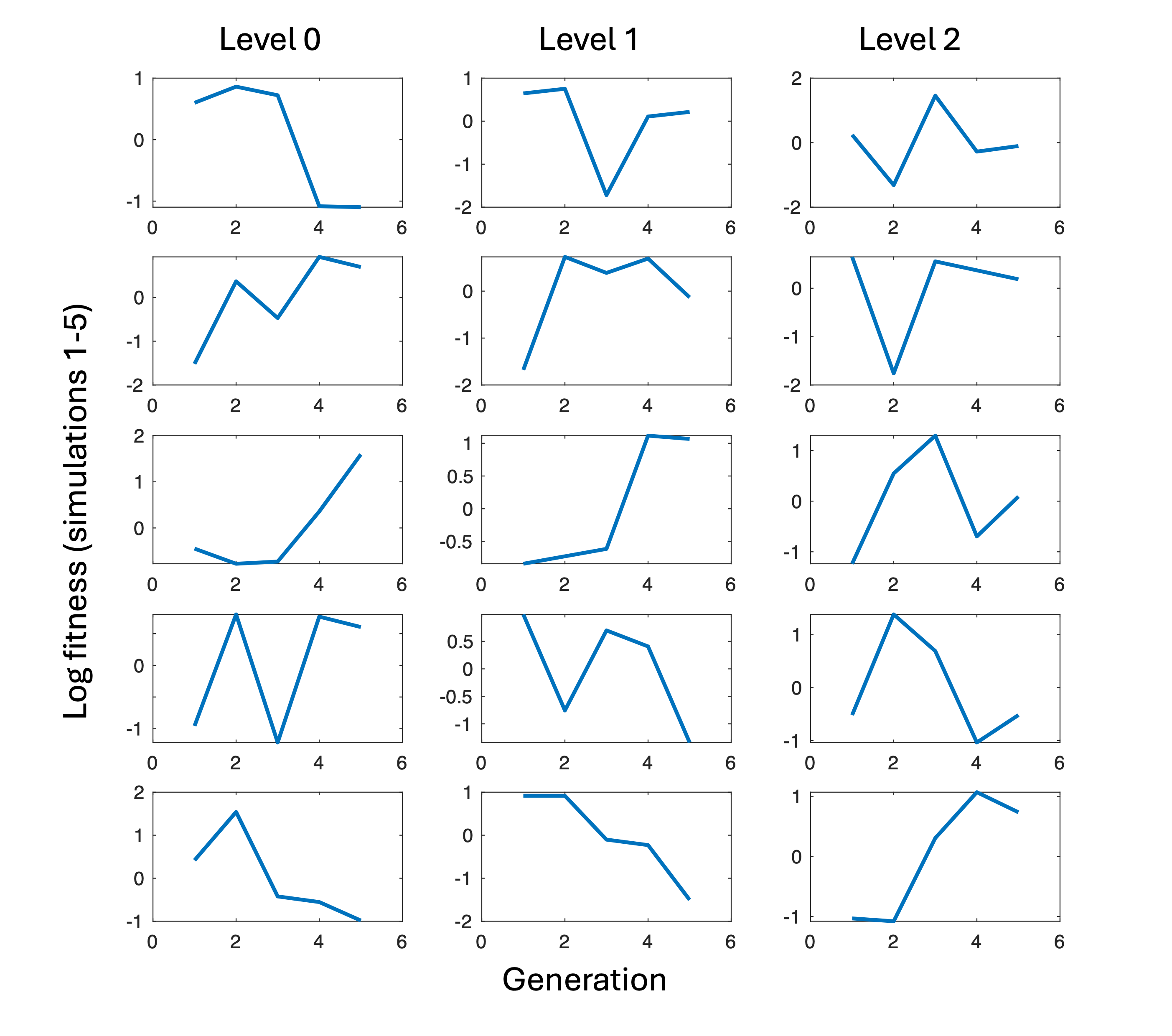}
\caption{Fitness across levels for 5 synthetic examples.  Mean value of $f_l$ is shown across all individuals or groups at level $l$.}
\label{fig:fig2}
\end{figure}

Fig. \ref{fig:fig2} illustrates the fitness profiles generated by the ground-truth simulations in our dataset, as specified above.  As shown, the simulations illustrate the potential for conflicts between levels, as suggested by the theoretical analysis of Sec. \ref{sec:theory}.  For instance, simulation 3 appears to show generally increasing fitness at all levels, while simulation 2 shows increasing fitness at levels 0 and 1 and generally decreasing fitness at level 2, while simulation 5 appears to show the opposite.

Table \ref{tab:perf} then shows the performance of the methods as described above when trained on this synthetic data.  As shown, using coalescent simulations appears to be beneficial for all methods.  In addition, we see for the VO and SPSA methods that the relative values of the Wasserstein distance metric are in general agreement with those of the Euclidian metric, suggesting that the former is an effective training metric for this problem.  While SPSA appears to slightly outperform VO on both metrics, the best performance overall (using the Euclidean metric to compare across all models) is achieved by MC-EM with coalescent simulations.  The results suggest that both variational and stochastic gradient-based methods should be considered and compared when fitting parameters of evolutionary models, and that coalescent methods are expected to be beneficial across methods.

%\subsection{Results on Synthetic Evolutionary Process}

\begin{table}[!t]
\caption{Comparing optimization approaches using forward and coalescent simulations.  Left half of table shows Wasserstein metric for VO and SPSA and ELBO for MC-EM (ELBO objective values are scaled by $10^{-3}$); right half shows Euclidean distance of estimated city positions to ground-truth for all methods, up to an abitrary rotation and translation.  Mean and standard deviation across 5 ground-truth samples are shown for all metrics. \newline}
    \label{tab:perf}
\scalebox{1}{
  \centering
  \begin{tabular}{ccccccc}%{llllll}
    \toprule
     & \multicolumn{3}{c}{Wasserstein distance / ELBO} & \multicolumn{3}{c}{Euclidean distance}  \\
    Simulator & VO  & SPSA  & MC-EM  & VO & SPSA & MC-EM  \\ \midrule
    Forward & $0.25\pm0.02$ & $0.24\pm0.01$ & $0.68\pm0.02$ & $0.81\pm0.24$ & $0.55\pm0.05$ & $0.97\pm0.28$ \\ 
    Coalescent & $0.23\pm0.03$ & $0.23\pm0.01$ & $0.53\pm0.21$ & $0.63\pm0.09$ & $0.44\pm0.01$ & $0.32\pm0.04$ \\ 
    \bottomrule
    \end{tabular}}
\end{table}

\section{Discussion}\label{sec:disc}

We have introduced.a general model of multilevel selection based on the Kantorovich Monad, which models the interaction of evolutionary units across arbitrarily many levels in a unified way, while depending only on a small number of key parameters.  We have shown through theoretical analysis that our model is capable of capturing antagonistic and cooperative effects between selection at all levels.  Further, we have provided a unified framework for optimizing the parameters of such multilevel evolutionary processes, and we have introduced a hierarchical Wasserstein objective and multilevel coalescent process, which we have shown can be used with both variational and gradient-based methods to provide efficient optimizers.  Finally, we consider below some of the directions for future work.

\noindent \textbf{Models of tumor evolution.}  A general problem in cencer genomics is to identify key driver variants, which may serve as the therapeutic targets.  Prior analyses have shown that the number of variants with non-neutral effects in cancer may be substantially larger than previously thought \cite{kumar2020passenger}, and methods have been proposed to identify driver genes and variants by inferring the evolutionary histories of individual tumors \cite{salichos2020estimating,yang2022lineage}.  The algorithmic methods we propose may be adapted to learn generative models of natural evolutionary processes, such as those in occuring in cancer, and identify key features such as driver variants and tumor clonal structure (see \cite{abecassis2021clonesig}).  Further, as discussed, it has been argued that concepts from multilevel selection theory may be applied to cancer \cite{okasha2006evolution}; we thus envisage that aspects of our multilevel model may be relevant in this context, although the framework is not intended to be a direct representation of a particular biological process. 

\noindent \textbf{Mathematical biology.}  More generally, our model may serve as a test-bed for analyzing multilevel biological systems and their properties.  For instance, it has been argued that microbial systems exhibit evolutionary dynamics at multiple levels \cite{frank2022microbial}, and multilevel selection has been proposed to play an important role in the major evolutionary transitions \cite{calcott2011major}.  By providing a general approach to fitting evolutionary models with arbitrary numbers of layers, our approach provides a potential method for identifying the number of evolutionary levels in a system \cite{simpson201110how,warrell2020cyclic}, as well as analysing the interaction and emergence of units at multiple levels in a tractable context \cite{krakauer2020information}.

\noindent \textbf{Optimization and Evolutionary Computation.}  Finally, our model and optimization methods may be applied directly to problems in evolutionary computation.  It has been shown that certain problem structures benefit from multilevel evolutionary search strategies in identifying solutions \cite{watson2006compositional,watson1999hierarchically}. Further, bi-level optimization methods have been shown to be effective for particular problem structures \cite{alesiani2023implicit}. Our framework provides a natural test-bed for exploring and generalizing such methods.  Our approach may also be viewed in the context of recent approaches to use category theory as a framework for deriving equivariant architectures and algorithms in a machine-learning \cite{gavranovic2024categorical}, where equivariance is achieved in our model through the use of multilevel Wasserstein distance functions.

\printbibliography

@inproceedings{alesiani2023implicit,
  title={Implicit bilevel optimization: differentiating through bilevel optimization programming},
  author={Alesiani, Francesco},
  booktitle={Proceedings of the AAAI Conference on Artificial Intelligence},
  volume={37},
  number={12},
  pages={14683--14691},
  year={2023}
}

@article{van2005metric,
  title={The metric monad for probabilistic nondeterminism},
  author={van Breugel, Franck},
  journal={Draft available at http://www.cse.yorku.ca/\textasciitilde franck/research/drafts/monad.pdf},
  year={2005},
  publisher={Citeseer}
}

@inproceedings{leordeanu2008smoothing,
  title={Smoothing-based optimization},
  author={Leordeanu, Marius and Hebert, Martial},
  booktitle={2008 IEEE Conference on Computer Vision and Pattern Recognition},
  pages={1--8},
  year={2008},
  organization={IEEE}
}

@article{rice2004mathematical,
  title={Evolutionary Theory: Mathematical and conceptual foundations},
  author={Rice, Sean H},
  year={2004},
  publisher={Springer}
}

@article{simpson201110how,
  title={How Many Levels Are There? How Insights from Evolutionary Transitions in Individuality Help Measure the Hierarchical Complexity of Life},
  author={Simpson, Carl},
  journal={Major Transitions in Evolution Revisited},
  pages={199},
  year={2011}
}

@book{okasha2006evolution,
  title={Evolution and the levels of selection},
  author={Okasha, Samir},
  year={2006},
  publisher={Clarendon Press}
}

@book{watson2006compositional,
  title={Compositional evolution: the impact of sex, symbiosis and modularity on the gradualist framework of evolution},
  author={Watson, Richard A},
  year={2006},
  publisher={Mit Press}
}

@article{frank2020generalized,
  title={The generalized Price equation: forces that change population statistics},
  author={Frank, Steven A and Godsoe, William},
  journal={Frontiers in Ecology and Evolution},
  volume={8},
  pages={240},
  year={2020},
  publisher={Frontiers Media SA}
}

@article{traulsen2006evolution,
  title={Evolution of cooperation by multilevel selection},
  author={Traulsen, Arne and Nowak, Martin A},
  journal={Proceedings of the National Academy of Sciences},
  volume={103},
  number={29},
  pages={10952--10955},
  year={2006},
  publisher={National Acad Sciences}
}

@article{tolstikhin2017wasserstein,
  title={Wasserstein auto-encoders},
  author={Tolstikhin, Ilya and Bousquet, Olivier and Gelly, Sylvain and Schoelkopf, Bernhard},
  journal={arXiv preprint arXiv:1711.01558},
  year={2017}
}

@inproceedings{moretti2021variational,
  title={Variational combinatorial sequential Monte Carlo methods for Bayesian phylogenetic inference},
  author={Moretti, Antonio Khalil and Zhang, Liyi and Naesseth, Christian A and Venner, Hadiah and Blei, David and Pe’er, Itsik},
  booktitle={Uncertainty in Artificial Intelligence},
  pages={971--981},
  year={2021},
  organization={PMLR}
}

@inproceedings{zhang2018variational,
  title={Variational Bayesian phylogenetic inference},
  author={Zhang, Cheng and Matsen IV, Frederick A},
  booktitle={International Conference on Learning Representations},
  year={2018}
}

@article{zhang2020improved,
  title={Improved variational Bayesian phylogenetic inference with normalizing flows},
  author={Zhang, Cheng},
  journal={Advances in neural information processing systems},
  volume={33},
  pages={18760--18771},
  year={2020}
}

@article{okasha2021cancer,
  title={Cancer and the levels of selection},
  author={Okasha, Samir},
  year={2021}
}

@book{calcott2011major,
  title={The major transitions in evolution revisited},
  author={Calcott, Brett and Sterelny, Kim},
  year={2011},
  publisher={MIT Press}
}

@inproceedings{watson1999hierarchically,
  title={Hierarchically consistent test problems for genetic algorithms},
  author={Watson, Richard A and Pollack, Jordan B},
  booktitle={Proceedings of the 1999 Congress on Evolutionary Computation-CEC99 (Cat. No. 99TH8406)},
  volume={2},
  pages={1406--1413},
  year={1999},
  organization={IEEE}
}

@article{warrell2020cyclic,
  title={Cyclic and multilevel causation in evolutionary processes},
  author={Warrell, Jonathan and Gerstein, Mark},
  journal={Biology \& Philosophy},
  volume={35},
  number={5},
  pages={50},
  year={2020},
  publisher={Springer}
}

@incollection{felsenstein2004inferring,
  title={Inferring phylogenies},
  author={Felsenstein, Joseph},
  booktitle={Inferring phylogenies},
  pages={664--664},
  year={2004}
}

@article{spall2001stochastic,
  title={Stochastic optimization, stochastic approximation and simulated annealing},
  author={Spall, James C},
  journal={Wiley Encyclopedia of Electrical and Electronics Engineering},
  year={2001},
  publisher={Wiley Online Library}
}

@article{ruth2024review,
  title={A review of Monte Carlo-based versions of the EM algorithm},
  author={Ruth, William},
  journal={arXiv preprint arXiv:2401.00945},
  year={2024}
}

@article{kumar2020passenger,
  title={Passenger mutations in more than 2,500 cancer genomes: overall molecular functional impact and consequences},
  author={Kumar, Sushant and Warrell, Jonathan and Li, Shantao and McGillivray, Patrick D and Meyerson, William and Salichos, Leonidas and Harmanci, Arif and Martinez-Fundichely, Alexander and Chan, Calvin WY and Nielsen, Morten Muhlig and others},
  journal={Cell},
  volume={180},
  number={5},
  pages={915--927},
  year={2020},
  publisher={Elsevier}
}

@article{salichos2020estimating,
  title={Estimating growth patterns and driver effects in tumor evolution from individual samples},
  author={Salichos, Leonidas and Meyerson, William and Warrell, Jonathan and Gerstein, Mark},
  journal={Nature communications},
  volume={11},
  number={1},
  pages={732},
  year={2020},
  publisher={Nature Publishing Group UK London}
}

@article{gavranovic2024categorical,
  title={Categorical deep learning: An algebraic theory of architectures},
  author={Gavranovi{\'c}, Bruno and Lessard, Paul and Dudzik, Andrew and von Glehn, Tamara and Ara{\'u}jo, Jo{\~a}o GM and Veli{\v{c}}kovi{\'c}, Petar},
  journal={arXiv preprint arXiv:2402.15332},
  year={2024}
}

@article{krakauer2020information,
  title={The information theory of individuality},
  author={Krakauer, David and Bertschinger, Nils and Olbrich, Eckehard and Flack, Jessica C and Ay, Nihat},
  journal={Theory in Biosciences},
  volume={139},
  pages={209--223},
  year={2020},
  publisher={Springer}
}

@article{abecassis2021clonesig,
  title={CloneSig can jointly infer intra-tumor heterogeneity and mutational signature activity in bulk tumor sequencing data},
  author={Ab{\'e}cassis, Judith and Reyal, Fabien and Vert, Jean-Philippe},
  journal={Nature communications},
  volume={12},
  number={1},
  pages={5352},
  year={2021},
  publisher={Nature Publishing Group UK London}
}

@book{frank2022microbial,
  title={Microbial life history: the fundamental forces of biological design},
  author={Frank, Steven A},
  year={2022},
  publisher={Princeton University Press}
}

@article{yang2022lineage,
  title={Lineage tracing reveals the phylodynamics, plasticity, and paths of tumor evolution},
  author={Yang, Dian and Jones, Matthew G and Naranjo, Santiago and Rideout, William M and Min, Kyung Hoi Joseph and Ho, Raymond and Wu, Wei and Replogle, Joseph M and Page, Jennifer L and Quinn, Jeffrey J and others},
  journal={Cell},
  volume={185},
  number={11},
  pages={1905--1923},
  year={2022},
  publisher={Elsevier}
}

%\bibliography{main}

\end{document}